\documentstyle[prl,aps,twocolumn,epsf]{revtex}

\newcommand{\doverd}[2]{\frac{\partial #1}{\partial #2}}

\begin{document}
\draft
\title{Relativistic dust disks and the Wilson-Mathews approach}

\author{
    Wilhelm Kley and Gerhard Sch\"afer}
\address{
    Theoretisch-Physikalisches Institut, Friedrich-Schiller-Universit\"at,\\
    Max-Wien-Platz 1, D-07743 Jena, Germany}



\date{\today}

\maketitle
\begin{abstract}
Treating problems in full general relativity is highly complex
and frequently approximate methods are employed to simplify the solution.
We present comparative solutions of a infinitesimally thin
relativistic, stationary, rigidly rotating disk obtained using
the full equations and the approximate approach suggested by Wilson \& Mathews. 
We find that the Wilson-Mathews method has about the same accuracy as the first
post-Newtonian approximation.
\end{abstract}
\pacs{95.30.Sf,04.25.Nx,04.25.Dm}
\section{Introduction}
The numerical solution of Einstein's field equations in three 
spatial dimensions in simulating for example the merging process of
two neutron stars is very complicated, and only very
recently have methods been developed for treating such problems in full general
relativity. One of the main difficulties is encountered in the treatment of
the radiation degrees of freedom.
An approximation scheme developed by Wilson \& Mathews \cite{wilson96}
simplifies the problem by imposing a conformally flat condition
on the three metric, throwing away all radiation degrees of freedom.
This reduces firstly the number of field equations and secondly reduces them
to Poisson like equations exclusively.
In general, the first post-Newtonian approximation of Einstein's
theory is identical to the
one of the Wilson-Mathews scheme, and in spherical
symmetry Wilson-Mathews is identical with Einstein.

As pointed out by Cook et al. \cite{cst96}, this approximation should
be viable and verifiable for systems in a true equilibrium state.
They investigated the accuracy of Wilson \& Mathews' approximative scheme on
isolated rapidly rotating relativistic stars,
and found largest errors of about 5\% for any quantity analyzed.
As the approximate scheme
has been used in calculating the evolution of a binary system
\cite{baum98} which
is very far from spherical symmetry, it is desirable to have test
cases, that depart stronger from spherical symmetry than rotating stars do.

Here we present the model of an infinitesimally thin, pressureless
disk that rotates rigidly. This problem was considered
by Bardeen \& Wagoner \cite{bw71} up to the tenth post-Newtonian
approximation and has been solved exactly by Neugebauer \& Meinel
\cite{dustdisk}. Thus, it serves as an ideal test bed to
check the validity of any numerical method solving the stationary,
axisymmetric field equations \cite{kley97} and in particular
Wilson \& Mathews' approximation scheme.
To demonstrate the general accuracy of the numerical method employed we
present the numerical solution of the full Einstein equations as well.
\section{Basic Equations}
To describe the stationary, axisymmetric equilibrium of an infinitesimally
thin disk we use a cylindrical coordinate system
($c=1$, $c$ velocity of light)
\begin{equation}
  x_0=t, x_1=\rho, x_2=z, x_3=\varphi
\end{equation}
where the disk is located in the equatorial plane (z=0), and the rotation axis
coincides with the $z$-axis.
\subsection{Full Einstein equations}
The general line element for an axisymmetric and stationary mass configuration
can be written as \cite{bw71}
\begin{equation}  \label{dsbw}
  ds^2 = e^{2 \mu} \left( d \rho^2 + dz^2 \right)
     + [\rho B e^{-\nu} ( d \varphi - \omega dt)]^2 - e^{2 \nu} dt^2,
\end{equation}
where the metric potentials $\nu, \mu, \omega$ and $B$ are functions
of $\rho$ and $z$.
As the disk has no vertical extension, the global solution of the
problem may be obtained by solving the vacuum field equations, where the
presence of the disk manifests itself through jump conditions
in the first (vertical) $z-$derivatives of the metric functions.
 
As shown by \cite{bw71} and \cite{kley97}, the vacuum Einstein equations
$G_{ij}=0$ for the metric (\ref{dsbw}) are then given explicitly by
\begin{eqnarray} \label{feld1} 
   \nabla(\nabla \nu) & = & \frac{1}{2} \rho^2 B^2 e^{-4 \nu}
          \nabla \omega \cdot \nabla \omega
        - \frac{1}{B} \nabla B \cdot \nabla \nu \\
   \frac{1}{\rho^2}  \nabla(\rho^2 \nabla \omega) & = & - \frac{1}{a}
          \nabla a \cdot \nabla \omega, \hspace{.40cm}
            \mbox{with} \hspace{0.25cm} a:= \frac{B^3}{e^{4\nu}} \label{feld2}\\
   \nabla(\rho\nabla B) & = & 0 \\
   \Delta_c \mu  & = & \frac{1}{4} \rho^2 B^2 e^{-4 \nu}
          \nabla \omega \cdot \nabla \omega
        + \frac{1}{B} \nabla B \cdot \nabla \nu \nonumber \\
     & - & \nabla {\nu} \cdot \nabla \nu  
     + \frac{1}{\rho} \doverd{\nu}{\rho}, \label{feld4}
\end{eqnarray}
with the $\nabla$-Operator in cylindrical coordinates ($\rho, z$), and
the Cartesian (in $\rho$ and $z$) Laplace operator $\Delta_c$.

The boundary conditions in the equatorial plane
within the disk region $\rho < \rho_d$, where $\rho_d$ denotes the outer
disk radius are then given by (see \cite{kley97})
\begin{eqnarray} 
   \nu_{,z} & = & 2 \pi \sigma \, 
          \frac{1+v_{\varphi}^2}{1-v_{\varphi}^2}  \label{bound1} \\
   \omega_{,z} & = & - 8 \pi \sigma \, 
      \frac{\Omega - \omega}{1 - v_{\varphi}^2} \\
   B_{,z} & = & 0 \\
   \mu_{,z} & = & -  2 \pi \sigma \label{bound4}.
\end{eqnarray}
Here $\sigma$ and $P$ denote the surface density and pressure, and
$ v_\varphi = \rho B e^{-2 \nu} (\Omega - \omega)$
is the orbital velocity of the gas.
We note that in case of a pressureless ($P=0$) dust disk
the metric function $B$ is identical to unity \cite{bw71}.

Radial hydrostatics yields the relation
$e^{\nu} \left( 1 - v_{\varphi}^2 \right)^{1/2} = const.$
in the disk, where $\rho \le \rho_d$. 
In the outer part of the equatorial plane, $\rho > \rho_d$, and along the
polar axis symmetry conditions are used for the metric coefficients.
At infinity $\nu, \mu,$ and $\omega$ tend to zero, and $B=1$.
\subsection{Wilson-Mathews formulation}
The method developed by Wilson and Mathews \cite{wilson96}
consists in approximating the line element 
(\ref{dsbw}) by using a spatially isotropic form and neglecting 
the non-isotropic parts. Specifically, for the problem at hand
we rewrite the exact metric as
\begin{eqnarray}  \label{wilson}
  ds^2 & = & e^{2\tilde{\mu}} \left[ \left( d\rho^2 + dz^2 \right)
     + \rho^2  ( d \varphi - \tilde{\omega} dt)^2 \right] \nonumber \\
    & + & \rho^2 \tilde{B}^2  ( d \varphi - \tilde{\omega} dt)^2 
    - e^{2 \tilde{\nu}} dt^2
\end{eqnarray}
with the functions $\tilde{\nu}$, $\tilde{B}$, $\tilde{\mu}$
and $\tilde{\omega}$.

By comparison with (\ref{dsbw}) we may identify
\begin{equation}
    e^{2 \mu} = e^{2 \tilde{\mu}}, \omega = \tilde{\omega}, \nu = \tilde{\nu}
\end{equation}
and for the metric function $\tilde{B}$ we obtain
\begin{equation}  \label{B}
    B^2 = (e^{2\tilde{\mu}} + \tilde{B}^2) e^{2\tilde{\nu}}.
\end{equation}
Applying the approximation by Wilson \& Mathews, we neglect the
non-isotropic part and set $\tilde{B}=0$. Simultaneously we have
to omit the corresponding field equation (for $B$), and we obtain the
approximate field equations from (\ref{feld1}), (\ref{feld2}),
and (\ref{feld4}) by setting $B=e^{\tilde{\nu} + \tilde{\mu}}$

\begin{eqnarray}
   \nabla(\nabla \tilde{\nu}) & = & 
         \frac{1}{2} \rho^2 e^{2(\tilde{\mu} -\tilde{\nu})}
          \nabla \tilde{\omega} \cdot \nabla \tilde{\omega}
        - \nabla (\tilde{\nu} + \tilde{\mu}) \cdot \nabla \tilde{\nu} \\
     \nabla(\rho^2 \nabla \tilde{\omega}) & = &
     - \rho^2 \nabla (3 \tilde{\mu} 
        - \tilde{\nu}) \cdot \nabla \tilde{\omega}, \\
   \Delta_c \tilde{\mu}  & = & 
       \frac{1}{4} \rho^2  e^{2(\tilde{\mu} -\tilde{\nu})}
          \nabla \tilde{\omega} \cdot \nabla \tilde{\omega}
          + \nabla \tilde{\mu} \cdot \nabla \tilde{\nu}
       + \frac{1}{\rho} \doverd{\tilde{\nu}}{\rho}
\end{eqnarray}
Furthermore, we have to substitute for $B$ in the equation for the
orbital velocity, and obtain
$v_\varphi = \rho  e^{(\tilde{\mu}- \tilde{\nu})} (\Omega-\tilde{\omega})$.
For our test case of a rigidly rotating dust disk
($B=1$) we may summarize the exact Einstein metric and the approximative 
Wilson-Mathews form as 
\begin{equation}  \label{dsein}
  ds^2 = e^{2 \mu} \left[ \left( d \rho^2 + dz^2 \right)
     + \rho^2  e^{-2(\nu+\mu)} ( d \varphi - \omega dt)^2 \right] \nonumber \\
      - e^{2 \nu} dt^2,
\end{equation}
and
\begin{equation}  \label{dswil}
  ds^2 = e^{2 \tilde{\mu}} \left[ \left( d \rho^2 + dz^2 \right)
     + \rho^2  ( d \varphi - \tilde{\omega} dt)^2 \right] \nonumber \\
      - e^{2 \tilde{\nu}} dt^2,
\end{equation}
respectively.
Comparing (\ref{dsein}) and (\ref{dswil}), it is striking that the
only difference between exact and approximate method for the dust disk example
lies in the function $e^{2\zeta} \equiv e^{2(\nu+\mu)}$.
The function $\zeta$ gives
the deviation from three-dimensional conformal flatness in Einstein's theory
and it has been shown that in
the first post-Newtonian approximation $\zeta$ vanishes for the dust disk
solution \cite{bw71}.
\subsection{Method of solution}
The field equations for the exact and the approximate case together
with the boundary conditions are solved
numerically for the problem of a rigidly rotating dust disk.
As the numerical procedure has been described in more detail
by Kley \cite{kley97} we give here only a brief sketch of the method.
To prevent problems with the conditions at infinity the computational
domain $[0:\infty; 0:\infty]$ is mapped onto the unit square where the radius
$\rho_d$ of the disk is located at $1/2$. 
This domain is covered here by $50 \times 50$ gridcells and the partial
differential equations are discretized to second order spatial accuracy on this
grid yielding a coupled system of equations. Together with the
boundary jump conditions these equations are then solved iteratively until
convergence has been achieved. 
The accuracy of the solutions may be estimated using the analytical solution
\cite{dustdisk}
and post-Newtonian approximation \cite{bw71} of the dust disk.
\section{Results}
To give an idea of the degree of non-isotropy of the three metric for
the exact Einsteinian disk we display in Figure 1 the behaviour of the
function $\zeta(\rho)$ within the disk for different values of the central
redshift. In the figure the numerically obtained solution of the
field equations (\ref{feld1})-(\ref{feld4}) together with the boundary
conditions (\ref{bound1})-(\ref{bound4}) is compared to the exact solution
as given by \cite{dustdisk}. The errors over the whole disk for all
central redshifts are usually well below 1\% and only at the outer edge
of the disk the worst deviation is about 2\% for $z=3$.
This demonstrates the accuracy
of the numerical method employed, even for the moderate gridsize used.
\begin{figure}
\epsfxsize=8.0cm
\epsfbox{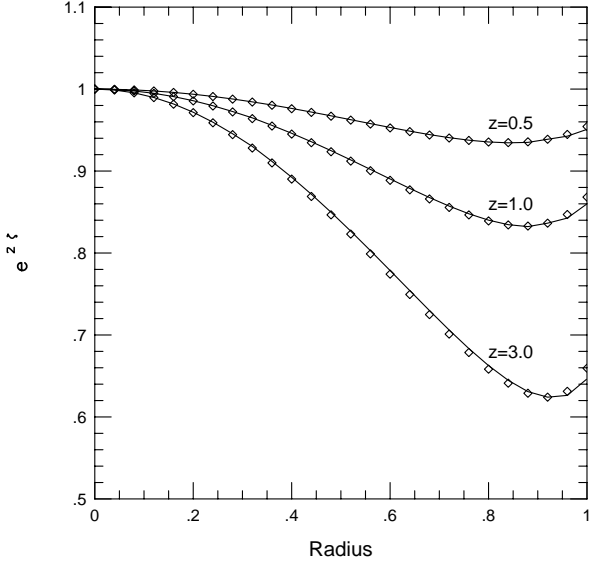}
  \caption{The function $e^{2 \zeta}$ which gives the deviation from local
  flatness plotted for the Einsteinian dust disk as function of the 
   normalized
    radius $\rho/\rho_d$ for three different central redshifts of the disk.
    The diamonds refer to the numerical solution of the full
   Einstein equations, while the solid lines refer to the exact solution.}
\end{figure}

\begin{figure}
\epsfxsize=8.0cm
\epsfbox{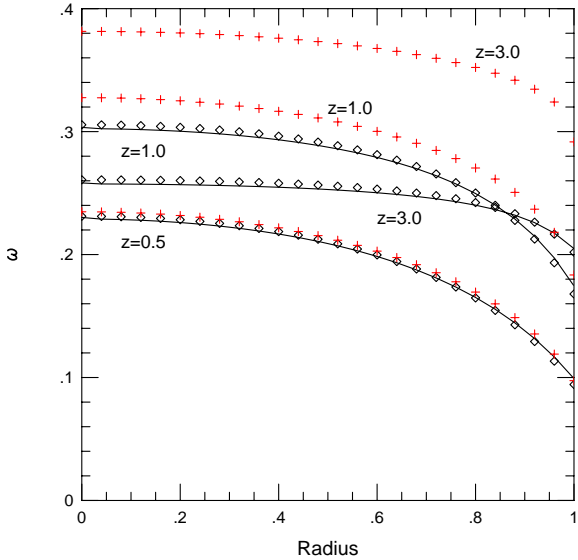}
  \caption{The metric function $\omega$ as a function of the normalized
    radius $\rho/\rho_d$ for three different central redshifts of the disk.
    The diamonds refer to the numerical solution of the full Einstein
    equations, while the crosses
    denote the numerical solution obtained by using the
    Wilson-Mathews approach. The solid lines refer to the exact solution.}
\end{figure}
We compared the numerically obtained approximate metric coefficients
$\tilde{\nu}$, $\tilde{\mu}$ and $\tilde{\omega}$ versus
radius with the exact solution and find that in case of the "lapse"
potential $\nu$ the deviation from the exact case is always
below 2.5\%, even for central redshifts as large as 3. For the function
$\mu$ the deviation is larger, 2.3\% for $z=0.5$, 5.5\% for $z=1.0$,
and about 20\% for $z=3$. The results for the metric function
$\omega$ are displayed in Fig.~2. The errors are very large for
higher redshifts, about 9\% for $z=1$ and up to 48\% for $z=3$.
The approximate solution (crosses) shows also the wrong trend of
the evolution of $\omega$ with 
redshift. While in the Einsteinian disk case $\omega$ increases first with
redshift, and then decreases for $z>2$, in the approximative solution 
$\omega$ seems to increase steadily with $z$.

\begin{figure}
\epsfxsize=8.0cm
\epsfbox{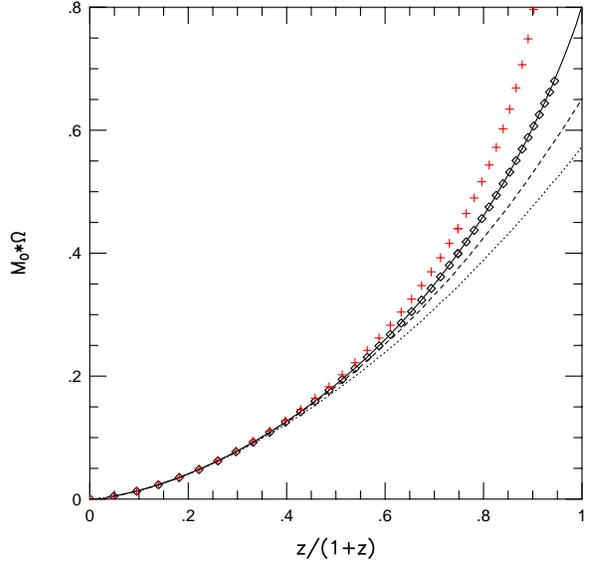}
\caption{
The quantity $\Omega M_0$ as a function of
central redshift $z$ for the dust disk. The open squares correspond to the
numerical solution of the full Einstein equations, the crosses to the
approximate Wilson-Mathews procedure.
The solid line refers to the exact solution,
and the dotted and dashed line to the first and second post-Newtonian
approximation respectively, of the Einstein equations.}
\end{figure}

The overall deviations of the metric functions from the exact case 
may be understood qualitatively by looking at the line elements (\ref{dsein})
and (\ref{dswil}). As the lapse is not too much affected by the simplification
in the spatial three metric it will be calculated quite accurately
in the Wilson-Mathews approach.
The rotational coefficient $\omega$ is affected mostly and hence shows
the largest errors.
We may nevertheless infer that at least for central redshifts up to $z=0.5$
the approximate solution is very accurate, but for higher
relativistic disks the agreement deteriorates. Note, that $z=.55$ was
the maximum (polar) redshift tested in the comparison for rotating stars
by Cook et al. \cite{cst96}.

As a further illustration of the range of applicability of the Wilson-Mathews
approximation, it is interesting to plot global quantities.
In Fig.~3 we display the dependence of the dimensionless quantity $\Omega M_0$
versus the parameter $\gamma=z/(1+z)$.
One notices again the very good agreement of the numerical and exact 
solution of the full Einstein equations. Interestingly, the
post-Newtonian approximations approach the exact solution from below 
while the Wilson-Mathews approximation lies always above the exact solution. 
In case of the full Einsteinian disk for infinite central redshift
$\Omega * M_0$ reaches a finite maximum of about $0.8$.
\begin{figure}
\epsfxsize=8.0cm
\epsfbox{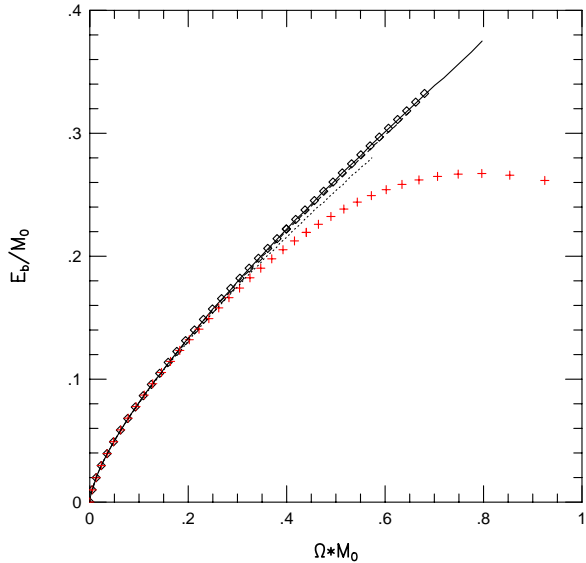}
\caption{
The binding energy as a function of $\Omega M_0$.
The symbols are identical to Figure 3.
}
\end{figure}
\begin{figure}
\epsfxsize=8.0cm
\epsfbox{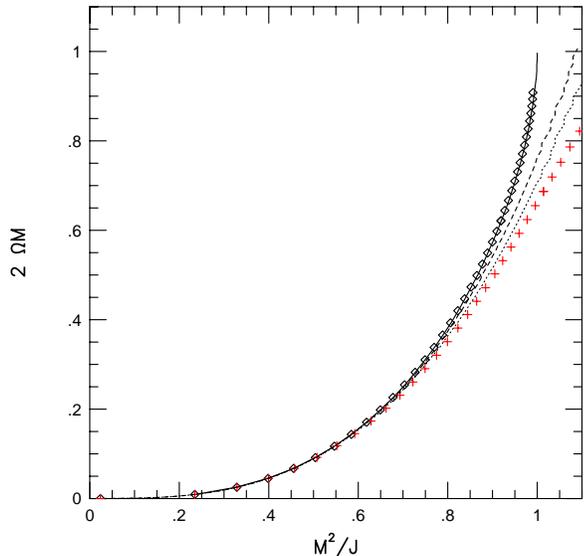}
\caption{
Dimensionless combination of the parameters measurable at infinity:
gravitational mass $M$, angular momentum $J$ and angular velocity $\Omega$.
The symbols are identical to Figure 3.
}
\end{figure}

In Fig.~4 the normalized binding energy $E_b/M_0=1-M/M_0$ is plotted
versus $\Omega M_0$. For a given rest mass $M_0$ and angular
rotation rate $\Omega$
the Wilson-Mathews approximation yields a lower binding energy
for the dust disk than the Einstein disk.
The dotted line for the post-Newtonian solution ends at $\Omega * M_0 = 0.57$
because this refers already to $z=\infty$ (cf. Fig.~3). 

Finally, in Figure 5 the dimensionless rotation rate $M \Omega$
is plotted versus the dimensionless total angular momentum $M^2/J$,
parameters that are measurable at infinity.
The value $2 \Omega M = M^2/J =1$ corresponds to the extreme Kerr
limit of the dust disk solution (see \cite{dustdisk}).
Again, the Wilson-Mathews approximation lies beyond the post-Newtonian
approximation and does not yield the correct black hole limit.
\section{Conclusions}
We have calculated numerically model sequences of a rigidly
rotating disk of dust using a) the full Einstein equations and
b) the modified Wilson-Mathews scheme.
By comparing those to the known exact solution by
Neugebauer \& Meinel \cite{dustdisk}, we have shown that the Wilson-Mathews
approximation yields results which are accurate to about 5\% if the
central redshift of the disk is smaller than 0.5; but may be significantly
higher for stronger relativistic models. We conclude that the
Wilson-Mathews approach for solving the Einstein equations
yields results which are at most as accurate as the first
post-Newtonian approximation, at least for rigidly rotating disks of dust.
Hence, its application to stronger relativistic configurations
e.g. in close, compact binary systems has to be handled with caution.

\end{document}